# Millimeter-Wave System for High Data Rate Indoor Communications

Lahatra Rakotondrainibe, Yvan Kokar, Gheorghe Zaharia, and Ghaïs El Zein
IETR-INSA, UMR CNRS 6164, 20, Av. des Buttes de Coësmes, CS 14315, 35043 Rennes Cedex, France
lrakoton@insa-rennes.fr; yvan.kokar2@insa-rennes.fr; gheorghe.zaharia@insa-rennes.fr; ghais.el-zein@insa-rennes

*Abstract*— **This paper presents the realization of a wireless Gigabit Ethernet communication system operating in the 60 GHz band. The system architecture uses a single carrier modulation. A differential encoded binary phase shift keying modulation and a differential demodulation scheme are adopted for the intermediate frequency blocks. The baseband blocks use Reed-Solomon RS (255, 239) coding and decoding for channel forward error correction (FEC). First results of bit error rate (BER) measurements at 875 Mbps, without channel coding, are presented for different antennas.**

## I. Introduction

One of the most promising solutions to achieve a gigabit class wireless link is to use millimeter-waves (MMW), especially at 60 GHz, for the carrier frequency. Aspects including global regulatory and standardization, justification of using 60 GHz bands, 60 GHz propagation and antennas and key system design issues were addressed in [1-2]. Recently, the IEEE 802.15.3c, ECMA and Very High Throughput (VHT) groups were formed to normalize the future wireless personal area networks (WPAN) systems, which operate in the 60 GHz band [3][4]. Hence, different architectures have been analyzed to develop new MMW communication systems for commercial applications. The choice of the architecture has a large impact on the system complexity and power consumption.

This paper describes the design and implementation of a 60 GHz system for high data rate wireless communications. The first system application in a point-to-point configuration is the high-speed files transfer. The system must operate in indoor, line-of-sight (LOS) domestic environments. Since the 60 GHz radio link operates reliably only within a single room, an optical fiber link is added to the intermediate frequency (IF) block in order to cover the whole house.

The structure of this paper is as follows. Section II describes the wireless communications system. In this section, IF and RF blocks are first presented. Then, the RS coder, the frame structure, the preamble detection and synchronization are described in baseband blocks. First measurement results are shown in Section III. Eye diagram and BER performance results are analyzed in this section. Conclusions and future work are drawn in the final section.

## II. Wireless communications system

The block schemes of the system transmitter and receiver are shown in Fig. 1 and Fig. 2 respectively. The Gigabit-Ethernet interface is used to connect a home server to a wireless link with around 800 Mbps bit rate. The system has a single carrier transmission scheme based on a differential encoded binary phase shift keying (DBPSK) modulation and a non-coherent demodulation. DBPSK modulation is used to avoid the phase ambiguity at the receiver. Compared to higher order constellations or multi-carrier modulation (OFDM), this system is more resistant to phase noise and power amplifiers (PAs) non-linearities. OFDM requires large back-off for PAs in the transmission, high stability and low phase noise for local oscillators which tend to be costly and reduce the power efficiency.

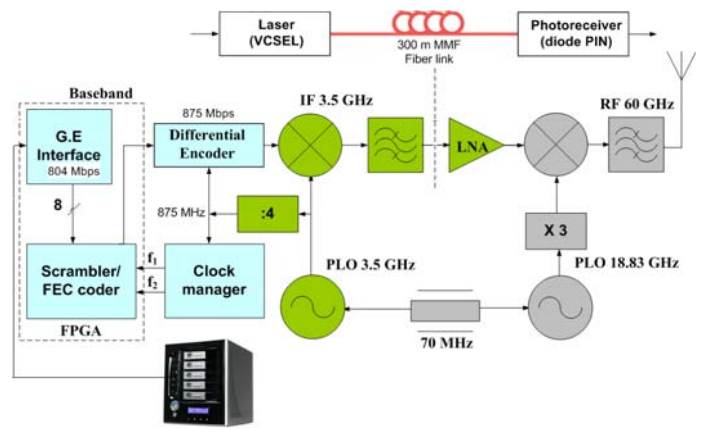

Figure 1.  60 GHz Wireless Gigabit Ethernet transmitter (Tx)

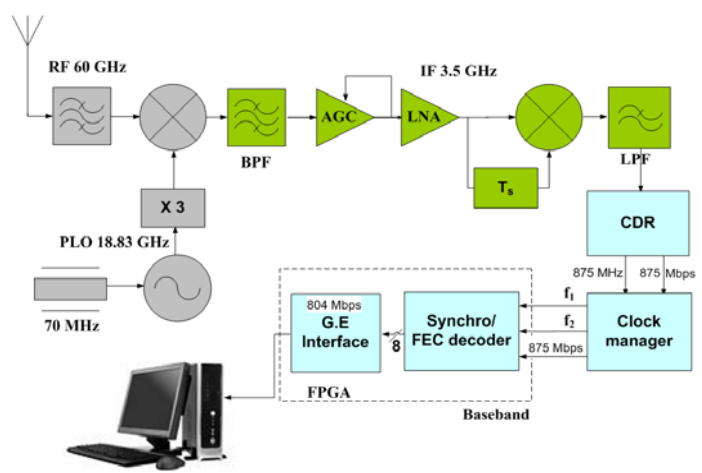

Figure 2.  60 GHz Wireless Gigabit Ethernet receiver (Rx)

## A. Intermediate and Radiofequency (RF) system architecture

At the transmitter, after the channel coding and the scrambling, the input data are differentially encoded using an XOR and a D flip-flop by PECL circuits. The differential encoder performs the modulo two addition of the input data bit with the previous output bit. The obtained data are used to modulate an IF carrier generated by a 3.5 GHz phase locked oscillator (PLO) with a 70 MHz external reference. The IF modulated signal is fed to a band-pass filter with a bandwidth of 2 GHz and transmitted to a 300 meters fiber link. The multimode fiber (MMF) with a bandwidth from 10 kHz to 10 GHz is enough to cover the frequency range of interest. This IF signal is used to modulate directly the courant of the Vertical Cavity Surface Emitting Laser (VCSEL) operating at 850 nm. The fiber input power does not exceed -3 dBm. The optical signal is converted to an electrical signal by a PIN diode ADOPCO. Following the fiber, an IF amplifier with a gain of 16 dB is used to offer sufficient power level at the input of the RF block. This block is composed of a mixer, a frequency tripler, a PLO at 18.83 GHz and a band-pass filter (59-61 GHz). The local oscillator frequency is obtained with an 18.83 GHz PLO with the same 70 MHz reference and a frequency tripler. The phase noise of the 18.83 GHz PLO signal is about –110 dBc/Hz at 10 kHz off-carrier. The upper sideband is selected using a band-pass filter. The 0 dBm obtained signal is fed to the horn antenna with a gain of 22.4 dBi and a half power beamwidth (HPBW) of 12°.

At the receiver, the RF signal at 60 GHz is fed to a band-pass filter to minimize the noise floor. The obtained signal is down-converted to the IF signal which is fed to a band-pass filter with a bandwidth of 2 GHz. An automatic gain control (AGC), with a dynamic range of 20 dB, is used to ensure a quasi-constant signal level at the demodulator input. A low noise amplifier (LNA) with a gain of 40 dB is used to achieve sufficient gain. A simple differential demodulation enables the coded signal to be demodulated and decoded. Compared to a coherent demodulation, this method is less performing in additive white Gaussian noise (AWGN) channel. However, the differential demodulation is more resistant to multi-path interference effect. Indeed, the significant impact on the system caused by the radio channel is the frequency selectivity which induces inter-symbol interference (ISI). The differential demodulation, based on a mixer and a delay line (Ts = 1.14 ns), compares the signal phase of two consecutive symbols. Following the loop, a low-pass filter (LPF) with 1 GHz frequency cut-off removes the high-frequency components of the obtained signal. For the clock and data recovery (CDR) circuit, long sequences of '0' or '1' must be avoided. Therefore, the use of a scrambler at the transmitter and a descrambler at the receiver is necessary.

## B. Baseband (BB) system architecture

Baseband blocks (BB-Tx & BB-Rx) are implemented in programmable circuit FPGA Xilinx Virtex 4. The BB-Tx is composed of the channel coding, the frame formatting block and the scrambler. The channel coding is realized with the RS (255, 239) coder, using byte data. A serial to parallel (S/P) conversion is necessary if the data source is a pattern generator, as shown in Fig. 3. The frame structure is shown in Fig. 4. The frame format consists of 4 preamble bytes, 239 data burst bytes, 1 frame header byte and 16 check bytes. To achieve the frame synchronization in packet-based communication systems, a known sequence (preamble) is sent at the beginning of each packet. The used preamble is a pseudo-noise (PN) sequence of 31 bits with one additional bit to provide 4 bytes. One more "frame header" byte should be added besides 4 preamble bytes to assure the scrambling operation, as it will be explained later.

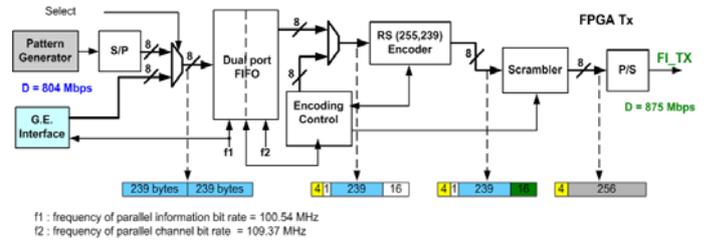

Figure 3. Transmitter baseband architecture

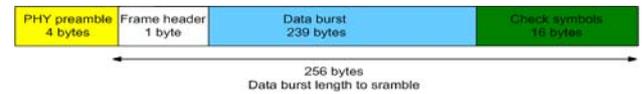

Figure 4. Frame structure

Due to the frame structure, two clock frequencies $F_1$ and $F_2$ are required, such as:

$$\frac{F_1}{F_2} = \frac{239}{260} \text{, or } F_2 = \frac{IF}{4} = 875 \text{ MHz} \rightarrow F_1 = 804.32 \text{ MHz}.$$

$$\text{Then, } f_1 = \frac{F_1}{8} = 100.54 \text{ MHz, and } f_2 = \frac{F_2}{8} = 109.37 \text{ MHz}.$$

Using the byte data, two frequencies $f_1$ and $f_2$ are required for the FPGA. In addition, the maximum frequency supported by the FPGA is limited (up to 200 MHz). To realise the frame format, data bytes from the S/P are written into the FIFO dual port at 100.54 MHz. The FIFO dual port has been set up to use two different frequencies for writing in at $f_1$ and reading out at $f_2$. Therefore, the reading can be started when the FIFO is half full. The encoding control is a module that monitors the RS encoder and the scrambler. The RS encoder reads one byte every clock period. After 239 clocks periods, the encoding control interrupts the bytes transfer during 21 triggered clock periods. During this interruption, 4 preamble bytes are added; moreover the encoder takes 239 data bytes and appends 16 control bytes to make a code word of 255 bytes. The encoding control adds 1 "frame header" byte for the balanced scrambling operation so that the number of 256 data coded bytes (except 4 preamble bytes) is a multiple of 8. The scrambler is realised with a sequence of 8 bytes (*i.e.,* a pseudorandom sequence of 63 bits + 1 bit). The scrambling facilitates the clock recovery at the receiver (eliminating long sequences of '0' or '1'). A scrambler of 8 bytes is chosen in order to provide a low cross-correlation with the preamble sequence (4 bytes). This method prevents false detections of the preamble within the scrambled data. The final step consists of a serial to parallel conversion of byte information before differential encoder interface.

Fig. 5 shows the receiver baseband architecture. At the receiver, recovered data from CDR at 875 Mbps are S/P converted into a byte frame.

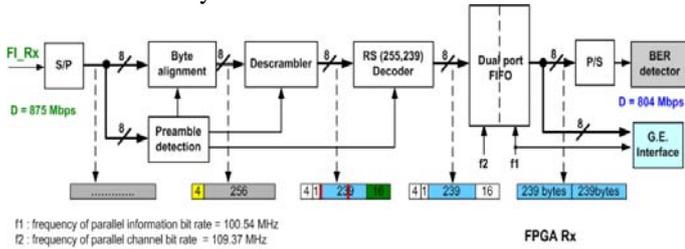

Figure 5.  Receiver baseband architecture

Fig. 6 presents the preamble detection architecture. The method is to perform a correlation of each received signal (32 bits) with the preamble sequence (4 bytes). In addition, each correlator must analyze a 1-bit shifted sequence. Therefore, the preamble detection is performed with 32 + 7 = 39 bits (+ 7 because of different possible shifts of a byte). Then, each value of the correlation between of the preamble and the received sequence data is compared to the threshold. The threshold is chosen in order to obtain the best trade-off between a high value of the detection probability and a very small value of the false detection probability.

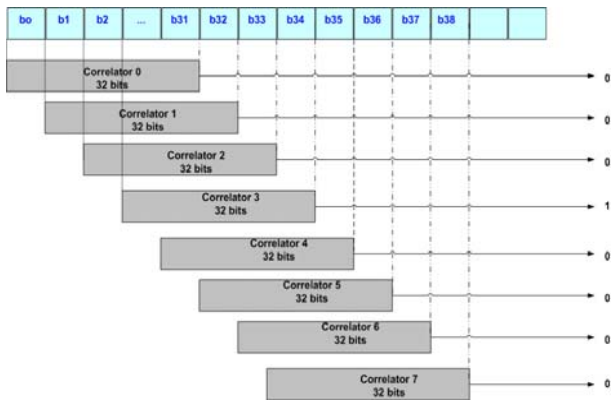

Figure 6.  Preamble detection architecture

In all, for the preamble detection, there are 8 corrrelators of 32 bits in a correlator-bank, as shown in Fig. 7.

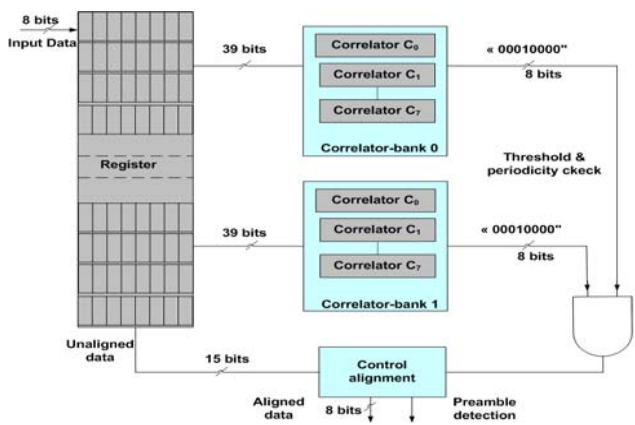

Figure 7.  Preamble detection with periodicity and control alignment

In order to obtain a very low false detection probability, the preamble periodical repetition is taking into account. The decision is performed from 264 successive bytes (preamble + data + preamble) stored in a register. If, in the both correlator-banks, the preamble detection is indicated by the same correlators $C_k$, this operation is validated. Therefore, with 8 + 7 = 15 successive bits, the byte alignment is established. If the preamble is detected, the descrambler performs a correlation of each received data (64 bits) with a same scrambler sequence (8 bytes). After the descrambling, the RS decoder processes and attempts to correct errors. The decoder can correct up to 8 bytes that contain errors. The decoder operates at higher rate 109.37 MHz. Therefore, data are written discontinuously in the FIFO dual port at frequency 109.37 MHz and another frequency 100.54 MHz read out continuously the data stored in a register. Finally, the reconstructed data at 804.32 Mbps are transmitted to the Gigabit Ethernet Interface or to the BER detector, as shown in Fig. 5.

III. MEASUREMENT RESULTS

Fig. 8 and Fig. 9 show measurement results of the frequency and impulse responses of RF blocks (Tx/Rx), including LOS channel. The goal was to measure the system bandwidth limitations and the effects of the multipath channel.

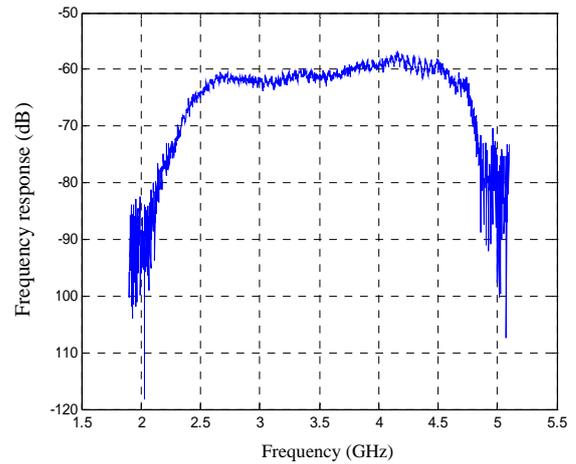

Figure 8.  Frequency response of RF (Tx-Rx) blocks (distance 10 m)

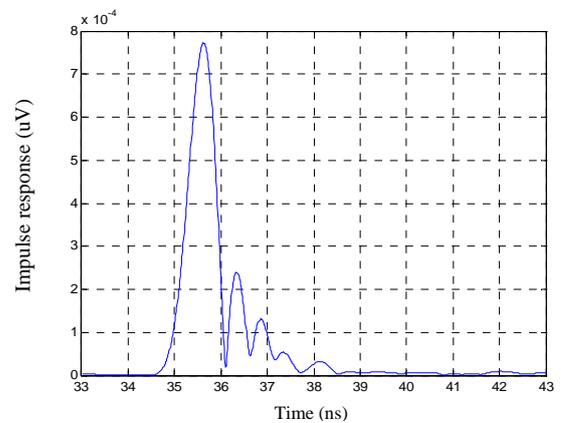

Figure 9.  Impulse response of RF (Tx-Rx) blocks (distance 10 m)

The measurements were performed in a corridor. After the calibration of the measurement set-up, a set of channel characteristics can be extracted: frequency and impulse responses, power attenuation and delay spread. As shown in Fig. 8 a frequency response of 2 GHz bandwidth is available. Measurements are performed using an HP 8753 D network analyzer. Side lobes in Fig. 9 are mainly due to RF components imperfections. Back-to-back tests have been realized and similar results were obtained. Directional antennas are essential in 60 GHz band to reduce the multi-path fading effects.

In order to validate the system performance, we used a HP70841B pattern generator to transmit PN sequences and a HP 708842B error detector at the receiver. However, the measurement was done without channel coding. As shown in Fig. 10, an open eye diagram was obtained at 10 m Tx-Rx distance, indicating a good transmission quality. From this eye pattern, the recovered data can be obtained by sampling at half period in order to be noise resistant. To determine the link quality, BER measurements according to the Tx-Rx distance are performed using an HP70842B error detector. Four antennas are used: two horns and two patch antennas. Each horn antenna has a 22.4 dBi gain and a 12° HPBW and each patch antenna has an 8 dBi gain and a 30° HPBW. Fig. 11 gives BER results at 875 Mbps, without channel coding.

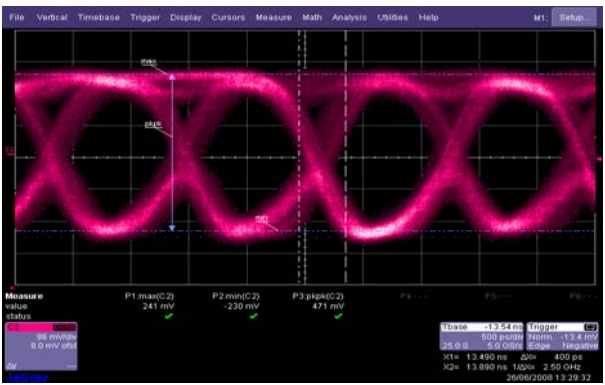

Figure 10. Eye diagram at 875 Mbps, point-to-point link, 10 m Tx-Rx distance (with horn antennas)

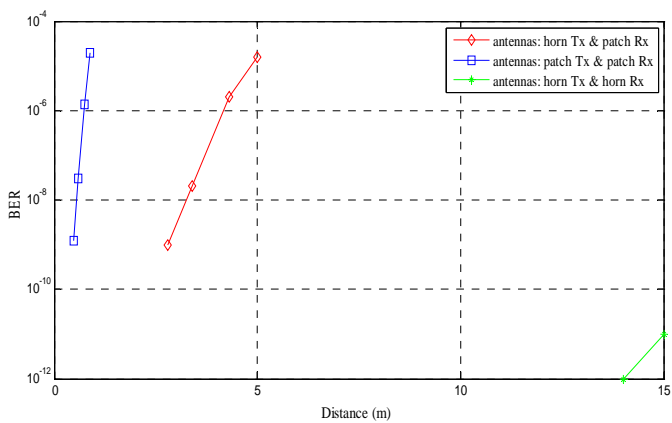

Figure 11. BER performance at 875 Mbps (without channel coding)

If high gain directive antennas are used, a very low BER can be achieved. This implies that with high-gain antennas and LOS conditions, very high data rates (several Gbps) can be obtained by using just a simple modulation scheme. System applications such as data transfer within a conference room (up to 20 m) can be used with high gain antennas. A drawback of this configuration that was experimentally observed is the human obstruction of the radio link [5]. The system using high gain antennas is acceptable for point-to-point communications links, with minimal multi-path interference. However, if antennas beamwidth is large, equalization should be used to overcome multi-path interference while maintaining a high data rate. If patch antennas are used, more gain is desired for RF front-ends to compensate the link budget. It was observed that with a large beamwidth, the influence of multipath propagation on BER results becomes critical for longer distances. Future work will provide the system performance results with channel coding.

CONCLUSION

This paper presents the design and the implementation of a 60 GHz system for high data rate wireless communications. The proposed system provides a good trade-off between performance and complexity. An original method used for the byte synchronization is also described. This method allows a high preamble detection probability and a very small false detection. BER measurement results were achieved to determine the link quality. For a reliable communication at data rates around 1 Gbps over distances up to 10 m, antennas must have a relatively high gain.

We planned to increase the data rate using higher order modulations. Equalization methods are still under study. The demonstrator will be further enhanced to prove the feasibility of wireless communications at data rates of several Gbps in different configurations, especially in the case of non line-of-sight (NLOS) environments.

ACKNOWLEDGMENT

This work is part of the research project Techim@ges supported by the French "Media & Network Cluster" and the COMIDOM project supported by the "Région Bretagne". The authors specially thank Guy Grunfelder (CNRS engineer) for his technical contributions during the system realization.